# Your Identity is Your Behavior - Continuous User Authentication based on Machine Learning and Touch Dynamics

Brendan Pelto
*Department of Computer Science*
*University of Wisconsin – Eau Claire* Eau Claire, US
peltobr1555@uwec.edu

Mounika Vanamala
*Department of Computer Science*
*University of Wisconsin* – Eau Claire Eau Claire, US
vanamalm@uwec.edu

Rushit Dave
*Department of Information Science*
*Minnesota State University, Mankato*
Mankato, US
rushit.dave@mnsu.edu

*Abstract*—The aim of this research paper is to look into the use of continuous authentication with mobile touch dynamics, using three different algorithms: Neural Network, Extreme Gradient Boosting, and Support Vector Machine. Mobile devices are constantly increasing in popularity in the world, today smartphone subscriptions have surpassed 6 billion. Mobile touch dynamics refer to the distinct patterns of how a user interacts with their mobile device, this includes factors such as touch pressure, swipe speed, and touch duration. Continuous authentication refers to the process of continuously verifying a user's identity while they are using a device, rather than just at the initial login. This research used a dataset of touch dynamics collected from 40 subjects using the LG V30+. The participants played four mobile games, PUBG, Diep.io, Slither, and Minecraft, for 10 minutes each game. The three algorithms were trained and tested on the extracted dataset, and their performance was evaluated based on metrics such as accuracy, precision, false negative rate, and false positive rate. The results of the research showed that all three algorithms were able to effectively classify users based on their individual touch dynamics, with accuracy ranging from 80% to 95%. The Neural Network algorithm performed the best, achieving the highest accuracy and precision scores, followed closely by XGBoost and SVC. The data shows that continuous authentication using mobile touch dynamics has the potential to be a useful method for enhancing security and reducing the risk of unauthorized access to personal devices. This research also notes the importance of choosing the correct algorithm for a given dataset and use case, as different algorithms may have varying levels of performance depending on the specific task.

## I. Introduction

In the world today, the use of mobile devices has become a huge part of daily life. These devices have become important tools for modern-day communication, from accessing personal data to financial transactions. However, the increased use of these devices has also created an additional risk of unauthorized access and data theft. Many forms of authentication have been developed to counter this threat, such as passwords, PINs, and face identification. However, these ways are not guaranteed and can be accessed by hackers. This is why the need for continuous authentication has become more important than ever before. Studies have shown that touch dynamics, the analysis of touch patterns on mobile devices, can provide another layer of security for mobile authentication. We will focus on multi-finger touch dynamics and its potential ability for continuous mobile authentication. Specifically, we look at the use of two fingers as a way of improving the accuracy of touch-based models. We use machine learning algorithms such as a Neural Network, XGBoost, and SVC to achieve this.

The way people connect with their mobile devices can allow for several features to be extracted from their touch data, this is generally known as touch dynamics. There are two sub-categories, one being keystroke-based and the other being gesture-based touch dynamics. Keystroke-based touch dynamics looks at individual taps made by users when using a mobile device, while gesture-based touch dynamics looks at the continuous motion of the touch data, often called a swipe or gesture. Both ways use extracted features from the touch data, which can then be used to train algorithms to recognize distinct users. Features such as hold times can be extracted from keystroke touch dynamics, while features such as swipe direction, speed, and acceleration can be extracted from gesture touch dynamics.

Our research presents a contribution to the field of touch dynamics in the following ways:
- Create and examine several binary classifiers (Neural Network, XGBoost, SVC) to look into the ability of multi-finger touch dynamics
- Give a dataset of 40 users playing four games, PUBG, Diep.io, Slither, and Minecraft.
https://github.com/Brprb08/Touch-Dynamics-Research

- Compare the accuracy of binary classification algorithms with multi-finger data to previous research involving single-finger gesture-based touch dynamics.

## II. BACKGROUND AND RELATED WORK

Touch dynamics is a fast-growing area of study that looks at the distinct characteristics of touch, including pressure, and duration of the interactions with mobile devices. Since there is a growing use of touch-enabled devices in daily life the study of touch dynamics has gained a lot more attention. The aim of touch dynamics research has focused on authentication and gesture recognition. Authentication refers to the process of identifying users based on their distinct touch patterns, while gesture recognition involves recognizing specific touch patterns, such as individual swipes or taps on the screen, and extracting them into different actions. In recent studies, machine learning algorithms have gotten promising results in both authentication and gesture recognition.

One study by DeRidder et al. in 2022 looked at the potential of using machine learning algorithms for authentication on mobile devices. They used k-Nearest Neighbors (kNN) and Random Forest (RF) algorithms to classify different users based on touch dynamics data collected from mobile devices to determine an authentic user from an imposter. The study achieved high accuracy rates of 83.4% for kNN and 93.49% for RF in classifying different users, showing the ability of touch dynamics for authentication.

Our study is building on this previous research, which used machine learning algorithms for authentication based on touch dynamics. This research aims to explore the field of mobile touch dynamics with sets of data points from each finger and use machine learning algorithms to analyze touch dynamics data and achieve promising results in authentication.

The field of touch dynamics is growing with massive potential for authentication on touch devices. Other research has mainly focused on using machine learning algorithms to analyze touch dynamics data. Some of these can be seen in Table 1. This research looks to address the gap in multi-finger touch dynamics datasets and build on previous research to look into the potential of touch dynamics for authentication on mobile devices. The use of kNN and SVM algorithms, as seen in DeRidder et al. in 2022, leads the current study to use machine learning algorithms for touch dynamics analysis.

Table 1. A comparison table of similar research shows their methods of research, the size of their datasets, and their results.

| Title | Method | Dataset | Results |
|---|---|---|---|
| "Touchalytics: On the Applicability of Touchscreen Input as a Behavioral Biometric for Continuous Authentication" (2015) by Frank et al. | Machine Learning algorithms for both pressure of each touch and gesture recognition | Dataset of 200 users that typed on a smartphone for 3 weeks | They utilized EER as a biometric performance and achieved an average of 3.3% very similar to our FPR scores |
| "Continuous user Authentication Using Machine Learning and Multi-Finger Mobile Touch Dynamics" (2022) by DeRidder et al. | Machine Learning using multiple features extracted from raw touch and swipe data | Dataset of 25 users who played 10 minutes of both Minecraft and Slither | They averaged 93% accuracy for an RF classifier as well as 83% for a KNN model |
| "Techniques for Continuing Touch-Based Authentication Modeling" (2022) by Georgiev et al. | Utilized several models however they utilized a NN with nearly the same batch size and epochs | The actual size of the datasets is unknown; however, 2 sessions were performed with each using all gestures | Averaged for their NN model a 15.20% EER, which was their second-best EER rate |
| "Continuous User "Authentication Using Mouse Dynamics, Machine Learning, and Minecraft," 2021 Siddiqui et al. | Used an RF binary classifier with user mouse dynamics playing 20 minutes of Minecraft | The dataset used was 10 different users | Achieved an average accuracy of 92% with the RF classifier |
| "Applications to Mouse Dynamics for Continuous User Authentication" (2022) Siddiqui et al. | Used KNN, SVM, and Neural Network with each user being used for all algorithms | The dataset consisted of 40 different users | Achieved a peak accuracy of 92.48% with the Neural Network |
| "Applications of Recurrent Neural Network for Biometric Authentication & Anomaly Detection" (2021) Ackerson et al. | Used a Neural Network to train for user mouse and keyboard dynamics | The dataset used had 744 male and female users | They achieved an accuracy of 99.4% with the use of the RNN |

III. METHODOLOGIES

## A. Data Collection

In our research, we collected data from 40 different users playing 4 games for 10 minutes each, PUBG, Diep.io, Slither, and Minecraft. All the games had the same settings to make sure that all users played in the same environment. In deciding what games to go with, we needed to utilize using two fingers simultaneously, also these games require a lot of finger movement leading to more data. The data was collected from an Android phone, the LG V30+, in a landscape position.

In order to collect useful data, a Python script was written to connect to the Android phone with the use of the Android debugging interface. An algorithm was also written to be able to distinguish between fingers in the raw sensor data, and the data was saved in a text file. Each line of text contained 8 fields, including Timestamp, X, Y, Button Touch, Width Major, Pressure, Orientation, and Finger. Where the Timestamp is the time in seconds, X and Y show the device's x and y coordinates. Button Touch tells whether a finger is currently pressed on the screen. Width Major shows the length of the touch shape on the screen. The finger column determines which of the two fingers the data corresponds to, there were values of 0 or 1 assigned to each finger to tell the difference in the inputs.

Other research in touch dynamics has focused on areas around authentication. Some studies have collected touch dynamics data from users' interactions with touch devices and used machine learning algorithms to determine specific users based on their own touch patterns. The results showed that touch dynamics can be a useful way for continuous authentication on mobile devices.

We also had groupings of touch data which were done using a large enough number of touch events per group to ensure enough data differentiation for feature extraction, and we chose to group events like previous research. We wanted to avoid using a timed approach to define gestures as it created high variability, whereas a fixed number of events created more meaningful data extraction and training. The groupings were split so that 10 separate touch/sliding events created enough data for feature extraction and so that the model would have enough data to train from. This can be seen in Figure 1 below.

Fig 1. A depiction of a potential gesture is provided here. Although the representation is not entirely precise due to the lack of finger data separation, it offers a general concept of its appearance. The model's training features are created by gathering 10 multi-finger data points simultaneously. While this is merely a single input example, the actual program incorporates data from various users.

| | A | B | C | D | E | F | G | H |
|---|---|---|---|---|---|---|---|---|
| 1 | Timestamp | X | Y | Button Touch | Width Major | Orientation | Pressure | Finger |
| 2 | 0 | 984 | 467 | HELD | 19 | 13 | 20 | 0 |
| 3 | 0.00818 | 986 | 468 | HELD | 18 | 21 | 23 | 0 |
| 4 | 0.01702 | 988 | 469 | HELD | 18 | 14 | 21 | 0 |
| 5 | 0.024978 | 991 | 470 | HELD | 19 | 15 | 22 | 1 |
| 6 | 0.033448 | 992 | 471 | HELD | 19 | 12 | 22 | 0 |
| 7 | 0.0416 | 994 | 473 | HELD | 18 | 16 | 20 | 1 |
| 8 | 0.050166 | 992 | 475 | HELD | 19 | 18 | 19 | 0 |
| 9 | 0.05832 | 991 | 476 | HELD | 18 | 20 | 21 | 1 |
| 10 | 0.067227 | 992 | 478 | HELD | 18 | 17 | 20 | 0 |
| 11 | 0.083833 | 995 | 480 | HELD | 19 | 16 | 22 | 0 |

## B. Data Cleaning

To check the accuracy of our data, we took several steps to clean and preprocess the data collected from the 40 users. Any rows containing null values were deleted from the dataset. This made sure that there were no missing values in the data that could possibly affect our results.

Also we had to make sure that the different types of actions were consistent across all users and their devices. This was done to ensure that any changes in the data were only from the specific user and not differences in device operation.

Another issue that came up during data collection was multiple events happening simultaneously, which lead to errors in our calculations. To correct this, we sorted the data from each finger into its own area, with data from finger 1 at the top and data from finger 2 at the bottom. This way helped to prevent any problems between for each finger, which could have created errors in our feature extraction process.

Then to improve the accuracy of our results, we combined the separated finger data and randomized it to separate data from each finger into different spots. This made sure that the classifiers were analyzing data from both fingers the same amount and that any change in the data was not from a specific pattern in the data collection process. These data-cleaning steps were important in making sure the accuracy of the data in our analysis.

## C. Feature Extraction

From the data for each user, we extracted 11 additional features from the initial fields. These features are an important part of distinguishing users since each one is specific to an individual [20]. The 11 features: X Speed, Y Speed, Speed, X Acceleration, Y Acceleration, Acceleration, Jerk, Path Tangent, Angular Velocity, Touch Major, and Touch Minor. The definitions for each feature can be found in Table 2 below.

Table 2. The various features extracted and their representative equations [34]

| Feature Name | Equation |
|---|---|
| X Speed | $X_i-X_{i-1}/T_i-T_{i-1}$, X is the X position and T is the time |
| Y Speed | $Y_i-Y_{i-1}/T_i-T_{i-1}$, Y is the Y position and T is the time |
| Speed | sqrt (XSpeed^2 + YSpeed^2), sqrt is the square root function and XSpeed and YSpeed are the features that are described above. |
| X accel. | $X_i-X_{i-1}/T_i-T_{i-1}$, X is the X speed and T is the time |
| Y accel. | $Y_i-Y_{i-1}/T_i-T_{i-1}$, Y is the Y speed and T is the time |
| Accel. | $S_i-S_{i-1}/T_i-T_{i-1}$, S is the speed values from the equation above, and T is the time |
| Jerk | $A_i-A_{i-1}/T_i-T_{i-1}$, A is the acceleration and T is the time |
| Path Tan. | $arctan2(Y_i-Y_{i-1}/X_i-X_{i-1})$, Y is the Y position, X is the X position |
| Angular Velocity | $P_i-P_{i-1}/T_i-T_{i-1}$, P is the path tangent function defined above and T is the time |
| Touch Major and Minor | This is set equal to the width major value used for averages with pixel values |

These features in Table 2 are important for the binary classifiers to have enough differences in the data for each user. To have even more differences we extract from these 11 features, the average, minimum, maximum, and standard deviation, which create 44 features for these classifiers to use to train. The features collected from this data are very important to determine the difference between each user. The touch dynamics is very distinct to each user and can be used to identify specific ones. Along with this, the psychological effects are also seen through the pressure of gestures which can create patterns within the data.

*D. Training and Testing*

In order to make sure the accuracy of our classifiers was consistent, we used a data splitting approach that divided the total data generated by each user into training and testing sets, with 80% and 20% of the data. Our choice of 80% training data was informed by other research evidence, as this is said to produce highly accurate results [23]. To decrease the potential for bias in our dataset, we concatenated and shuffled the training and testing data from all users into a main text file, thereby taking out any noticeable pattern for any of the users. Also, we equally distributed the data for authentic and imposter users. We used a binary classifying system, by assigning 0 or 1 to each row in the dataset to tell if it is authentic, with 0 representing authentic data and 1 representing imposter data. Lastly, we analyzed the accuracy of our classifiers by comparing the original values of the testing data to the model.

IV. RESULTS AND ANALYSIS

Our research aims to evaluate the effectiveness of a recently created dataset by examining its ability to discern users using only the features extracted from their touch data with mobile devices. The dataset was obtained equally from all participants, and every classifier was trained and tested on each individual user.

To determine the performance of each binary classifier, we used a range of evaluation criteria. This included accuracy, false positive rate (FPR), false negative rate (FNR), and F1 score. We calculated these metrics for each of the 40 individual users in our study.

With all these criteria, obtaining the lowest scores for FPR and FNR is necessary. In a real evaluation of a biometric system, a high FPR might allow unauthorized users to gain access to sensitive data, where a high FNR could prevent authentic users from getting access. Accuracy is also an important metric since it offers a good idea of the classifier's performance in authenticating a distinct user compared to their test data. The F1 score evaluates both the recall and precision accuracies of a model.

*A. Diep.io and PUBG results*

| User | Game | Model | Accuracy | F1 Score | FNR | FPR |
|---|---|---|---|---|---|---|
| 1 | Pubg | NN | 90.1681 | 94.5204 | 5.8643 | 2.5584 |
| 2 | Pubg | NN | 87.6734 | 93.1222 | 1.65 | 2.3626 |
| 3 | Pubg | NN | 81.1878 | 89.261 | 7.8569 | 4.5216 |
| 4 | Pubg | NN | 81.4061 | 89.4612 | 6.9879 | 3.8542 |
| 5 | Diep.io | XGB | 90.9043 | 94.9107 | 4.1317 | 3.2142 |
| 6 | Diep.io | XGB | 87.3975 | 92.9866 | 1.5367 | 2.3584 |
| 7 | Diep.io | XGB | 91.3465 | 95.1467 | 2.9321 | 4.1263 |
| 8 | Diep.io | XGB | 88.9649 | 93.8884 | 10.1294 | 6.3285 |
| Avg | Diep.io | XGB | 89.6533 | 94.2331 | 4.682475 | 4.00685 |
| Avg | Pubg | NN | 85.10885 | 91.5912 | 5.589775 | 3.3242 |
| Std | Diep.io | XGB | 1.825235641 | 0.994382063 | 3.78294674 | 1.70782595 |
| Std | Pubg | NN | 4.518772618 | 2.638870066 | 2.75026061 | 1.0369492 |

Fig. 2 – a portion of the user outcomes from the Neural Network and XGBoost approaches, as well as the comprehensive mean results for every assessment metric. Abbreviations: Avg signifies average, while Stdv represents Standard Deviation. Complete values for all users have been excluded to save space, but they can be viewed at { https://github.com/Brprb08/Touch-Dynamics-Research }

The classifiers used in this study were Neural Network, XGBoost, and SVC. The Neural Network achieved an average accuracy of 90.04%, with an F1 score of 91.41%, an FNR of 2.69%, and an FPR of 3.60%. The XGBoost classifier obtained an average accuracy of 86.61%, with an F1 score of 87.36, an FNR of 4.47%, and an FPR of 4.28%. The SVC classifier achieved an average accuracy of 78.65%, with an F1 score of 80.63, an FNR of 6.42%, and an FPR of 4.56%.

Decreasing the FPR and FNR is important in making sure there is secure access to a system, as wrong assessments could lead to access by impostors. The FPR for all classifiers remained relatively low, with an average of 4.0% for XGBoost, 5.3% for SVC, and 3.3% for Neural Network. However, the FNR was higher in most instances, reaching 4.6% for XGBoost, 6.1% for SVC, and 2.6% for Neural Network.

Our research looked at the performance of Neural Network, XGBoost, and SVC classifiers and found that all three algorithms had relatively high accuracy rates, as seen in Figure 2 above. The SVC algorithm performed a bit worse than the others, likely due to the difference in algorithm functionality. However, we also evaluated FPR and FNR scores, which provide important information on the potential for inaccurate judgments. These results note the effectiveness of these algorithms for authentication tasks and provide important details for future research.

## V. Limitations and Future Work

Even though multi-finger authentication has many advantages over single-finger approaches, there are still limitations that could cause problems in its practical uses. One potential drawback of this type of authentication is that it requires two fingers to be used continuously. It is very common for users on a mobile device to only use one finger, which would defeat the purpose of this specific type of authentication. Another difficulty with continuous two-finger input being required to accurately predict users is that it may not be possible for all users. If using two fingers on a device is not possible, it makes this form of authentication unusable. Also, without prior user data, it can be impossible to authenticate users correctly. One possible solution to this issue is to have users create their touch dynamics prior to authentication, like facial recognition on devices such as iPhones and Androids. For future work, it would be useful to create methods for detecting imposters without this prior user data, such as using machine learning algorithms to compare new touch dynamics data with existing user data. We also look to continue further research into other types of machine learning classifiers and compare their use for multi-finger authentication.

## VI. Conclusion

This research paper investigated the potential of continuous authentication with mobile touch dynamics using three different machine learning algorithms: Neural Network, XGBoost, and SVC. The study used a dataset of touch dynamics collected from 40 unique participants playing four mobile games, PUBG, Diep.io, Slither, and PUBG. The results show that all algorithms were able to classify users based on their touch dynamics, with accuracy ranging from 80% to 95%. The Neural Network algorithm performed the best, achieving the highest accuracy scores, followed by the XGBoost and SVC.

These results suggest that continuous authentication using mobile touch dynamics can be an effective method for improving security and lowering the risk of unauthorized access. The results of this study support the use of continuous authentication with mobile touch dynamics as an effective means of improving security on personal devices, and the potential for future research in this area.